Sub-micron Cu(In,Ga)Se$_2$ solar cell with efficiency of 18.2% enabled by a hole transport layer

Taowen Wang[1*], Longfei Song[2], Saeed Bayat[1], Michele Melchiorre[1], Nathalie Valle[2], Adrian-Marie Philippe[2], Emmanuel Defay[2], Sebastjan Glinsek[2], and Susanne Siebentritt[1*]

1. Laboratory for Photovoltaics (LPV), Department of Physics and Materials Science, University of Luxembourg, 41 rue du Brill, L-4422, Belvaux, Luxembourg.
2. Luxembourg Institute of Science and Technology (LIST), 41 rue du Brill, L-4422, Belvaux, Luxembourg

*Corresponding author: wangtaowenscu@hotmail.com , susanne.siebentritt@uni.lu

**Abstract**

Reducing the thickness of Cu(In,Ga)Se$_2$ solar cells is a key objective in order to reduce production cost and to improve sustainability. The major challenge for sub-micron Cu(In,Ga)Se$_2$ cells is the recombination at the backside. In standard Cu(In,Ga)Se$_2$ backside recombination is suppressed by a bandgap gradient, acting as a back surface field. This gradient is difficult to maintain in sub-micron thick absorbers. In this study, a hole transport layer passivates the back contact and enables efficient sub-micron Cu(In,Ga)Se$_2$ solar cells without the need of a Ga gradient. The backside passivation by the hole transport layer is as effective as an optimized Ga gradient, resulting in a significant increase in open-circuit voltage by 80 mV in comparison to the reference sample without passivation. Moreover, the hole transport layer exhibits good transport properties, leading to a fill factor as high as 77%. Photoluminescence quantum yield of 0.15% and solar cell efficiency above 18% are demonstrated in sub-micron Cu(In,Ga)Se$_2$ absorbers.

**Introduction**

Cu(In,Ga)Se$_2$ (CIGS) based thin film solar cells represent one of the photovoltaic (PV) technologies with highest efficiencies of 23.6% and proven stability[1-4]. However, these record CIGS solar cells are typically achieved using a rather thick absorber layer of ~2 μm thickness. While this thickness allows for high efficiency, it also demands a larger quantity of raw metals, such as In and Ga, resulting in higher production and environmental cost. Reducing the thickness of the absorber layer presents a compelling opportunity to decrease material consumption. However, the efficiency of these solar cells cannot be compromised since the cost of electricity generation is directly influenced by the efficiency. Higher efficiency translates to lower system costs, making it a key factor in driving cost-effectiveness of PV electricity.

Reducing the absorber thickness places challenges for the open circuit voltage, as well as for the short circuit current. The reduction in short circuit current is mainly due to reduced absorption in thinner layers. Experiments and simulations show that the absorption loss becomes severe only for absorbers of 500nm and below[5]. Absorbers that thin require advanced light management techniques[6-9]. However, in the range from 500 to 1000 nm absorber thickness, the efficiency is reduced mostly because of a significant decrease in open-circuit voltage $V_{oc}$ due to backside recombination[10,11]. The back contact interface between the molybdenum layer and the CIGS absorber exhibits a high recombination velocity exceeding $10^5$ cm/s[12-14]. To address backside recombination in standard CIGS solar cells, a Ga gradient is employed, introducing a conduction band increase towards the backside, which drives minority carriers away from the back contact, thereby reducing backside recombination, similar to a back surface field. Simulations have shown that the efficiency of sub-micron CIGS solar cells can be above 20% with a well passivated back surface[15]. In our previous study[16], we showed by experiment and simulation a significant improvement in $V_{oc}$ by reducing backside recombination for $CuInSe_2$ solar cells with a Ga backside gradient even in thick absorbers. However, when dealing with thinner absorbers, the limited deposition duration and reduced thickness of absorber pose challenges in establishing the necessary Ga gradient. To mitigate backside recombination experimentally, extensive experimentation has demonstrated the necessity of a dedicated and intricate 3-stage process to grow a CIGS absorber with a sufficiently steep conduction band gradient towards the back contact [17-20]. By optimizing the Ga gradient, the best efficiency, so far, of 17.5% has been achieved using 900 nm CIGS with an anti-reflection coating[19]. However, there are some issues with the required steep Ga gradient towards the backside of sub-micron CIGS solar cells. Firstly, the exact shape of the Ga gradient has a significant impact on the performance of solar cells[21], necessitating highly reproducible processes with minimal tolerance for fluctuations. This requirement may hinder large-scale production. Secondly, the region with high Ga/(Ga+In) ratio has been observed to exhibit a high recombination rate, resulting in a low minority carrier lifetime of approximately 100 ps[22], in line with the observation of deep defects for Ga/Ga+In ratios $> 0.5$[23]. Moreover, the bandgap gradient limits the effective absorption of low-energy photons due to the reduced thickness of the band gap minimum region[24]. Consequently, this leads to higher losses of low-energy photons in $J_{sc}$. In addition, the necessary steep bandgap gradient results in higher radiative $V_{oc}$ losses [25-27].

To overcome these challenges, implementing a hole selective transport layer (HTL) is an effective solution. The HTL should exhibit high resistance to minority carriers (electrons) and low resistance to majority carriers (holes). This function can be achieved through, for example, asymmetric conductivity and band offsets[28,29]. By employing this configuration, the minority carrier density dependent surface recombination can be minimized, and majority carriers can be freely transported. The concept of utilizing a HTL as a replacement for the Ga gradient holds significant promise. However, identifying a suitable material that can withstand the harsh growth conditions of CIGS, such as high temperature and selenium (Se) pressure, has been a challenge.

Recently[30], we successfully developed a functional HTL with good backside passivation and efficient hole transport properties. This HTL is prepared as a $CuGaSe_2$ layer covered by solution combustion synthesis prepared $In_2O_3$, which during the absorber growth process converts into $CuInSe_2$/$GaO_x$. Solution combustion synthesis is practical for laboratory-scale or wafer type production. However, since it involves spin coating it cannot be scaled to the m$^2$ sized substrates of thin film PV industry. Therefore, we additionally explore sputtered $In_2O_3$ in this study. We demonstrate that sputtered $In_2O_3$ matches the performance of solution processed $In_2O_3$ in terms of passivation and hole transportation, resulting in similarly improved $V_{oc}$ and fill factor (FF), promising prospects for large-scale application of this novel hole transport layer. Applying this HTL to sub-micron Cu(In,Ga)Se$_2$ enhances $V_{oc}$ by up to 80 mV while maintaining a FF of 77%. Further enhancement through rubidium fluoride RbF post-deposition treatment (PDT) yields a good photoluminescence (PL) quantum yield $Y_{PL}$ of 0.15% and an efficiency of 18.2% (active area) for sub-micron CIGS solar cells covered by an anti-reflection coating (ARC). An efficiency of 16.2% (full-area) has been certified. The difference can be attributed to non-optimised contact grids area, which blocks lights and results in loss of $J_{sc}$.

**Studied samples**

The samples analyzed in this study are briefly introduced here, with detailed sample preparation procedures discussed in the Methods section. The general sample structure is glass-Mo-(HTL-)CIGS-CdS-ZnO-ZnO:Al-Ni/Al grids(MgF$_2$), see **Fig. 6a**. Reference samples, denoted as MoRe, consist of CIGS directly grown on Mo without HTL. The HTL is grown as a CuGaSe$_2$/In$_2$O$_3$ stack deposited on Mo prior to absorber deposition. This stack transforms into Cu(In,Ga)Se$_2$/GaO$_x$ during absorber deposition, as elaborated below. The thickness of the original CuGaSe$_2$ layer ranges from around 100 nm to 200 nm, while In$_2$O$_3$ thickness varies from 10 nm to 40 nm. The CuGaSe$_2$ layer is prepared via co-evaporation at a substrate setting temperature of 356 °C, with a Cu/Ga ratio of 0.9-0.95 determined by energy-dispersive X-ray spectroscopy (EDS). In$_2$O$_3$ deposition is achieved through either solution combustion synthesis or RF sputtering. Prior to absorber deposition, this stack (Glass/Mo/CGS/In$_2$O$_3$) undergoes annealing at a substrate setting temperature of 500 °C for 20 minutes under vacuum of ~5×10$^{-9}$ Torr, with or without a ~40 nm layer of Cu on top. Cu annealing enhances the hole transport through the HTL, as discussed later. Although, we believe that the CIGS part of the HTL cannot contribute to $J_{sc}$ because the photo-generated electrons cannot pass through the GaO$_x$ due to the high conduction band offset, we address the concern about the actual absorber thickness, by adding the thickness of CIGS in the HTL to the actual absorber thickness. The total thickness is written as "thickness of CIGS in HTL + CIGS absorber thickness". The CIGS absorber for the record sub-micron (0.10+0.75) μm solar cells with an active area efficiency of 18.2% is prepared using a 3-stage process[31,32], with the 1$^{st}$ stage substrate setting temperature at 500 °C and 580 °C for the 2$^{nd}$ and 3$^{rd}$ stages. For all other CIGS absorbers, the 1$^{st}$ stage substrate temperature is 356 °C, with subsequent stages at 580 °C. The higher temperature in the 1$^{st}$ stage aims to enhance grain size and improve absorber quality. RbF PDT, if applied, is carried out immediately after the absorber growth without vacuum interruption. It's done at a substrate setting temperature of 280 °C under a Se flux of ~2.5×10$^{-6}$ Torr for 10 minutes, with RbF source temperatures of 450 °C. All discussed samples are coated with chemical bath deposited CdS, approximately 50 nm thick. Prior to CdS deposition, samples undergo etching in 5% KCN aqueous solution for 30 s; RbF-treated samples are additionally etched in 1.5 M NH$_4$OH for 3 minutes after KCN etching. To complete solar cell devices, an intrinsic zinc oxide/aluminum-doped zinc oxide stack is RF sputtered as a window layer, and a grid of Ni/Al is evaporated as a front contact. The record solar cell is covered with 90 nm MgF$_2$ as an ARC. Some samples are treated by heat light soaking: absorbers covered with the CdS buffer undergo treatment in a N$_2$ atmosphere at a substrate setting temperature of 80 °C, with an equivalent illumination intensity of 0.5 Sun and a duration of 3 hours.

**Ion exchange between In and Ga in the HTL**

In a previous study[30], the functionality of this HTL has been demonstrated for CuInSe$_2$ solar cells without Ga in the absorber with a bandgap around 1.0 eV. The HTL exhibits thermal stability, remaining physically at the backside despite the harsh growth conditions of the absorber. On the other hand, complete ion exchange between In and Ga occurs, converting CuGaSe$_2$/In$_2$O$_3$ into CuInSe$_2$/GaO$_x$. In the present study, we reaffirm the thermal stability of the HTL and the ion exchange process in the presence of Ga containing absorbers with somewhat wider band gap. We analyse a sample in detail after the absorber process, which is made from a solution deposited In$_2$O$_3$ layer with Cu annealing. Cross-section images from scanning transmission electron microscopy in **Fig. 1a** clearly depict the individual layers between the Mo back contact and the absorber. These layers are further distinguishable through EDS mapping of the same area. The top part of the back contact is identified as MoSe$_2$, rich in Mo and Se as shown in **Fig. 1b, c**, consistent with previous reports[33,34]. The subsequent layer is identified as Cu(In,Ga)Se$_2$, as Cu, In, Ga, and Se are detected (**Fig. 1d, e, f, c**). Additionally, a 30~40 nm thick layer is observed to be rich in O and Ga, with minimal In and Se content, strongly suggesting an exchange of In and Ga, leading to GaO$_x$ formation from the original In$_2$O$_3$, as observed in prior work[30]. However, unlike the case of pure CIS absorbers, where the CGS beneath the oxide layer completely converts into CIS after absorber deposition[30], in the case of CIGS absorbers, Ga is still present beneath the oxide layer. The Ga concentration, measured as Ga/(Ga+In) ] ratio, is comparable to that of the absorber layer above the oxide layer, as can be seen from the Ga and In profiles in **Fig. 1h**. Diffusion is primarily driven by temperature and concentration differences, the process appears to reach equilibrium when the Ga contents on both sides of the oxide are equal. Nonetheless, whether the bottom layer of the HTL is CuInSe$_2$ or Cu(In,Ga)Se$_2$, no discernible differences in passivation are observed, as demonstrated by the performance of the ensuing solar cells, discussed below.

Furthermore, the same Ga content between the CIGS layers on both sides of the GaO$_x$ could indicate as well that there is no Ga gradient toward the backside of the samples. However, the absorber thickness studied in TEM-EDS mapping (**Fig. 1e, f**) or line scans (**Fig. 1h**) is insufficient to confirm the absence of a Ga gradient over the whole depth of the absorber. To address this concern, secondary ion mass spectrometry (SIMS) is conducted on the same sample as the one studied by TEM. As the system is not calibrated for CIGS (since no standard sample has been available), the non-absolute SIMS data is presented in **Fig. 1i**, where each element distribution is normalized to its maximum within the measured range. Although the depth resolution of the low-energy SIMS used is a few nanometers, it increases to a few tens of nanometers due to the roughness of the different interfaces in the stack. Thus, individual CIGS and $GaO_x$ layers cannot be distinguished at the backside, but the interface between the absorber and $GaO_x$ can be located where the oxygen signal is half of its maximum. The region marked in blue in **Fig. 1i** can be identified with the HTL containing $GaO_x$/CIGS (from left to right). The absorber is to the left of the blue region, the Mo back contact is to the right. It is evident that the Ga and In compositions remain constant over most of the absorber depth, in particular towards the backside, indicating the absence of a Ga gradient that would passivate the back contact. Therefore, any observed passivation in our samples is solely attributed to the HTL rather than a Ga gradient. However, a small Ga gradient towards the frontside persists. Although the frontside Ga gradient may help mitigate front surface recombination[35], it becomes unnecessary when adequate passivation is achieved through methods such as proper band alignment[36-38] and heavy alkali PDT[1,38,39]. Consequently, a completely homogeneous absorber would improve collection of long-wavelength photons and reduce radiative losses in $V_{oc}$[25].

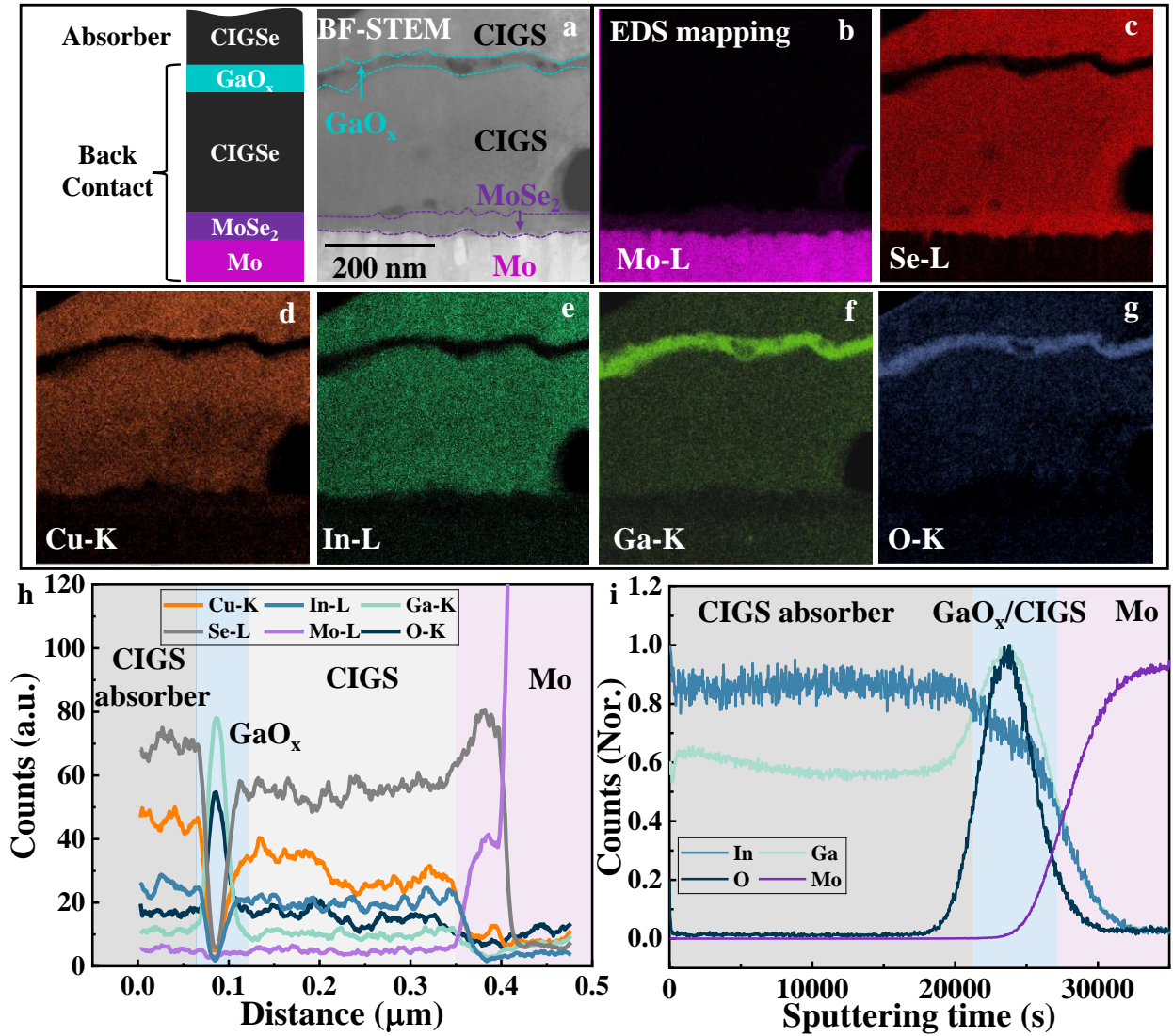

**Fig. 1| a**, Bright field (BF) scanning transmission electron micrograph (STEM). From bottom to top, the back contact region consist of Mo, MoSe$_2$, CIGS and GaO$_x$, in which the CIGS and GaO$_x$ are formed from CuGaSe$_2$/In$_2$O$_3$ via ion exchange between Ga and In; The EDS mapping of the same as the scanning transmission electron micrograph: **b**, Mo; **c**, Se; **d**, Cu; **e**, In; **f**, Ga; **g**, O. **h,** EDS line scans across the back contact interfaces. **i,** SIMS depth profiles show that there is no Ga gradient in the absorber towards the backside.

We attribute the flat Ga profile towards the backside to two main factors. Firstly, establishing a suitable Ga gradient within a constrained absorber thickness requires an optimized 3-stage process, where the Ga/In flux ratio during the 1$^{st}$ stage is significantly higher than that of the 3$^{rd}$ stage[15]. However, in our approach, we maintain the same Ga/In flux ratio in both the 1$^{st}$ and 3$^{rd}$ stages.

Secondly, in the case of the Cu-annealed sample, absorber growth starts with Cu-rich CIGS rather than $(In,Ga)_2Se_3$ precursors. This variation influences the reaction rate of Ga and In compounds with Cu during the 2$^{nd}$ stage, ultimately resulting in a flat Ga distribution towards the backside. As has been demonstrated in the past, when Cu is present throughout the absorber process, the Ga profile becomes much flatter and is only controlled by the Ga/In flux ratios during the time of the process[1,21,40].

**Good passivation but hole transport blocking**

The as-grown HTL consists of approximately 100 nm $CuGaSe_2$ covered by approximately 40 nm solution deposited $In_2O_3$. The layers after the absorber process are clearly visible in SEM cross-section images (see **Extended Data Fig. 1**). To evaluate the passivating effect of the HTL for sub-micron CIGS, time resolved PL (TRPL) analysis is performed on two sub-micron CIGS films deposited on HTL and directly on Mo, both grown in a classical 3-stage process without a Cu-annealing step at the beginning (**Fig. 2a**). The two absorbers are grown in the same co-evaporation process and the CdS front passivation layer is deposited in the same chemical bath process. The decay is best described as bi-exponential. The longer-term lifetime $\tau_2$ obtained from the 2-exponential fitting is used in the following as minority carrier lifetime, as it is more sensitive to backside recombination and therefore provides a better indication of passivation effects[41]. As illustrated in **Fig. 2a**, the lifetime of the HTL passivated sample improved significantly from 19 to 161 ns compared to the reference without HTL. Given that these samples are prepared by the same absorber and CdS processes, it is reasonable to infer that they have similar bulk and front surface lifetimes. Therefore, the longer $\tau_2$ indicates effective backside passivation, meaning a reduction in backside recombination. Further evidence for the passivation is provided by absolute photoluminescence spectroscopy, which allows to extract the quasi-Fermi level splitting $\Delta E_F$ from a fit to Planck's generalised law and separately the non-radiative $\Delta E_F$ loss from the PL quantum yield $Y_{PL}$ [$k_bT \times \ln(Y_{PL})$][26,42,43] **Fig. 2b,c** show that the HTL increases $\Delta E_F$ by about 80 meV, the same amount by which the non-radiative loss is reduced . However, when investigating the ensuing solar cells (Fig. 2 c,d) we detect two problems: the increase in $V_{OC}$ is lower than the increase in $\Delta E_F$ and the FF is very poor. The difference between $\Delta E_F$ and $V_{OC}$ indicates a gradient in at least one of the quasi Fermi levels, which occurs at or near the contacts[44]. We attribute this additional loss to hole transport blocking in the HTL, which restricts the forward diode current, resulting in a linear behavior of the *J-V* curve, as depicted in **Fig. 2c**. Consequently, this leads to an extremely low FF of approximately 30%, compared to the reference sample's FF of approximately 76%. This unfavorable blocking is primarily attributed to the high valence band offset between $GaO_x$ and the absorber, as previously discussed[30]. We had also proposed that the hole transport properties of $GaO_x$ are closely linked to the amount of Cu present. Cu can introduce deep defects near the valence band maximum of CIGS, assisting hole transport[45,46]. To address the issue of hole transport, here, we experiment with reducing the thickness of $In_2O_3$ and introducing an additional Cu annealing process to the oxide layer, aiming to enhance its hole transport properties.

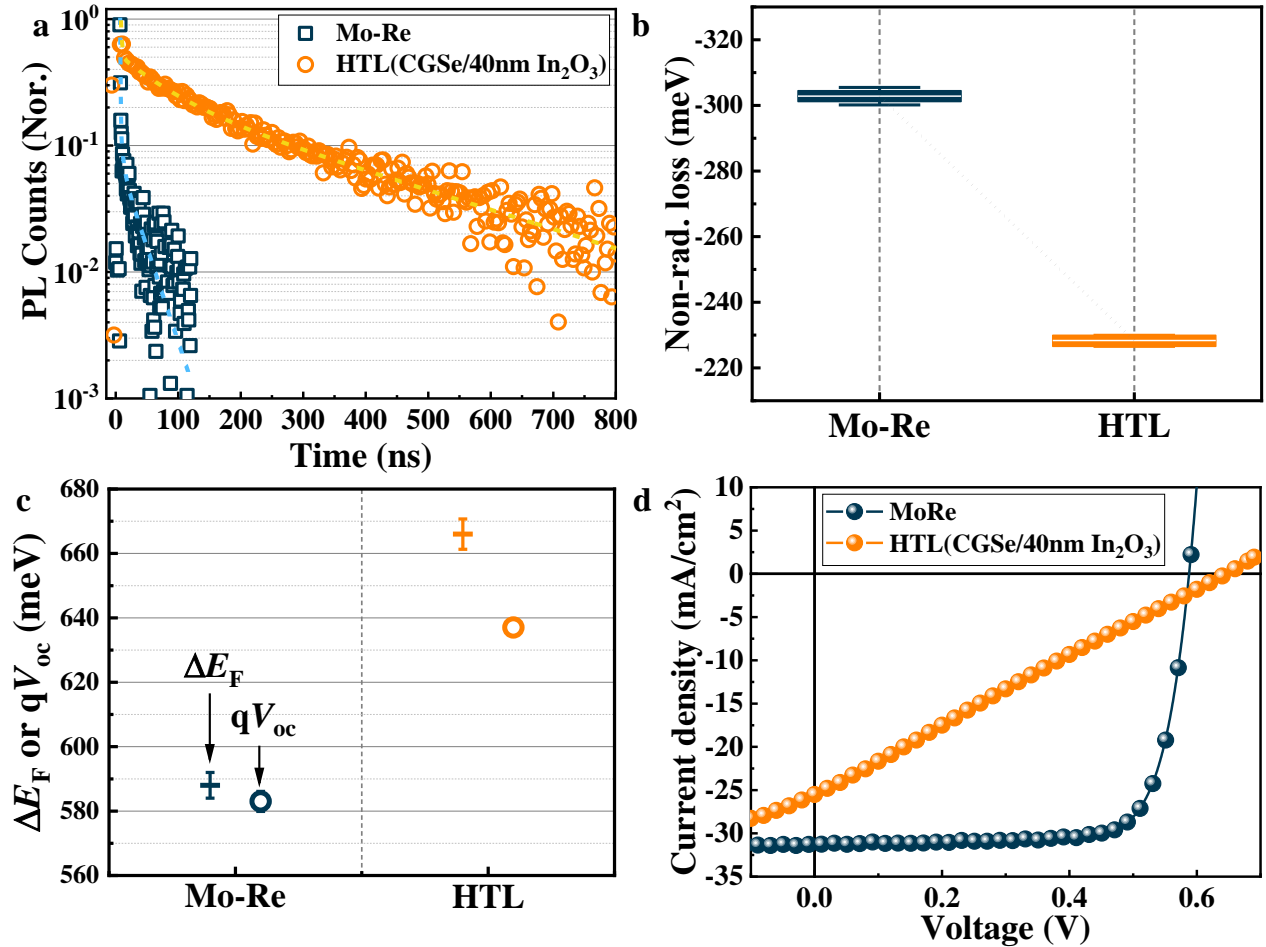

**Fig. 2| a,** TRPL measurements of reference sample and sample passivated by HTL made from 40 nm solution processed $In_2O_3$. The good passivation of the HTL improves the lifetime of the sample from ~19 to 161 ns. **b,** Non-radiative loss in $\Delta E_F$ of the two samples, which is determined by $k_bT \times \ln(Y_{PL})$. **c,** Comparison of $\Delta E_F$ and $qV_{oc}$ of these samples. Due to hole transport blocking caused by the HTL, the limited forward diode current results in an additional $V_{oc}$ loss, $\Delta E_F - qV_{oc}$.

**d,** Current density vs. voltage characteristics under illumination, indicating an extremely low FF of the HTL passivated sample due to blocking of hole transport.

**Reducing the thickness of In$_2$O$_3$**

Previous studies indicate that reducing the thickness of oxide layers, like Al$_2$O$_3$[47-49] and TiO$_2$[50,51], has demonstrated an improvement in hole transport, leading to increased fill factor (FF), but usually with a trade-off in passivation and consequently reduced $V_{oc}$. This trade-off between FF and $V_{oc}$ generally hinders the achievement of high-efficiency solar cells. Consistent with these findings, our study also reveals a similar trade-off (see **Extended Data Fig. 1a, b**). Decreasing the thickness of In$_2$O$_3$ from 40 nm to 20 nm and 10 nm progressively enhances the FF of the solar cells. Notably, solar cells with 10 nm In$_2$O$_3$ exhibit nearly identical FF (~77%) compared to the reference sample, indicating successful removal of hole transport blocking. However, the non-radiative $\Delta E_F$ loss of samples with thinner In$_2$O$_3$ are 20~30 meV higher than those with thicker In$_2$O$_3$, suggesting that excessively thin oxide layers fail to provide adequate backside passivation.

**Cu annealing: making the HTL conductive for holes**

To address the trade-off between passivation and hole transport, we introduce additional Cu annealing prior to the absorber deposition. This Cu annealing is considered because our previous work[30] has demonstrated that utilizing a Cu-rich CuGaSe$_2$ layer enables hole transport, suggesting that extra Cu may play a crucial role in hole transport. Furthermore, it has been observed that Cu can introduce deep defects in GaO$_x$, which may facilitate hole transport: This hole transport has been demonstrated in perovskite solar cells through GaO$_x$[45] or Ga$_x$In$_{2-x}$O$_3$[46] doped with Cu. The In$_2$O$_3$, from which the GaO$_x$ is formed during absorber deposition, is prepared either by solution combustion synthesis or by RF sputtering, with thicknesses of 20, 30, and 40 nm for solution deposited In$_2$O$_3$, and 30, 35, and 40 nm for sputtered In$_2$O$_3$. In all cases, approximately 40 nm of Cu is evaporated onto the surface of In$_2$O$_3$ at a substrate setting temperature of 200 °C. Subsequently, the temperature is rapidly increased from 200 to 500 °C at a rate of 50 °C/minute, followed by annealing under vacuum (~5×10$^{-9}$ Torr) at 500 °C for 20 minutes. Afterward, the substrate is cooled to 356 °C before the deposition of CIGS via a 3-stage process. The thickness of the absorbers deposited onto In$_2$O$_3$ prepared by solution combustion synthesis or RF sputtering is approximately (0.20+0.90) μm and (0.10+0.75) μm, respectively, determined through SEM cross-section imaging (see **Extended Data Fig. 1**). We remind here that the actual absorber thickness is 0.90 or 0.75 μm, the first number in the sum gives the thickness of the $Cu(In,Ga)Se_2$ layer underneath the oxide layer. As indicated in the literature, a CIGS thickness around 1 μm is barely sufficient to avoid $J_{sc}$ loss due to non-absorption[6-9,52-54]. Hence, the (0.20+0.90) μm and (0.10+0.75) μm CIGS absorbers may exhibit some non-absorption loss of $J_{sc}$, as will be discussed further.

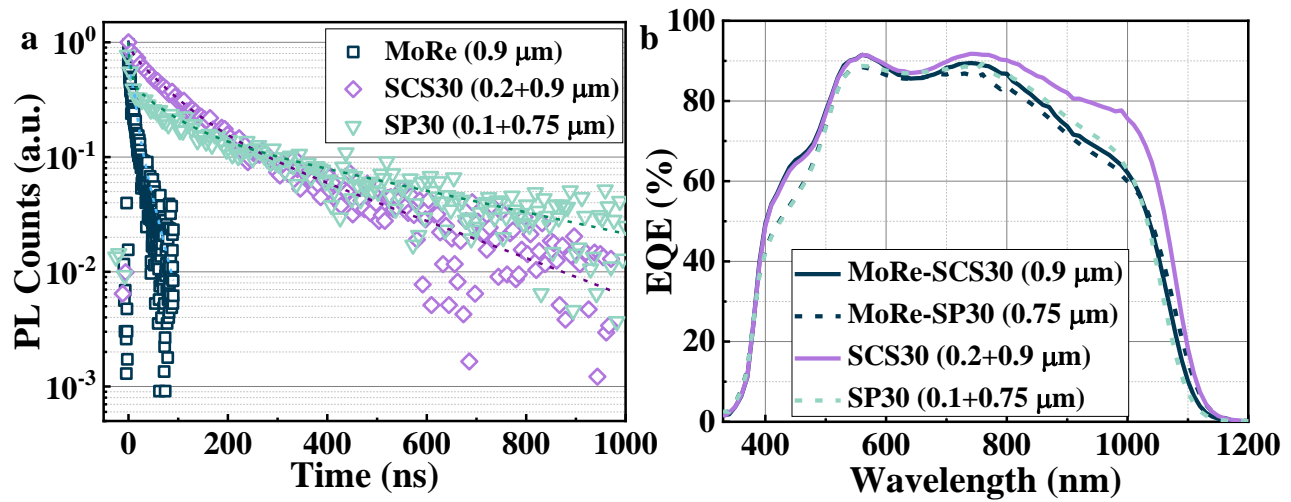

**Fig. 3| a,** TRPL measurements of a reference sample without HTL and samples passivated by a Cu-annealed HTL, prepared from a solution deposited or sputtered $In_2O_3$ layer with a thickness of 30 nm, labeled as SCS30 and SP30, respectively. With the HTL, the lifetime significantly improved from 19 ns to 164 ns and 264 ns for the SCS30 and SP30 samples, respectively, confirming the effective passivation achieved by either HTL. **b,** The EQE spectra of solar cells passivated by Cu-annealed HTL, together with their respective unpassivated references, demonstrate enhanced collection of long-wavelength photons, further supporting good backside passivation provided by the Cu-annealed HTL.

To evaluate the passivation of Cu-annealed HTL, TRPL analysis is conducted (**Fig. 3a**). Both solution and sputter-prepared $In_2O_3$, each with a thickness of 30 nm and labeled as "SCS30" and "SP30" respectively, exhibit a notable enhancement in lifetime. Specifically, when compared to the reference sample, the lifetime $\tau_2$ increased from 19 ns to 164 ns and 264 ns for the SCS30 and SP30 samples, respectively. The fast initial decay observed in the MoRe reference sample and the passivated SP30 sample can be due to many causes (surface recombination, potential fluctuations or drift in space charge regions formed at grain boundaries or at the surface). The substantial improvement in the lifetime $\tau_2$ confirms the effectiveness of the passivation. Furthermore, as shown by the EQE spectra (**Fig. 3b**), the good backside passivation is demonstrated by the better collection of photogenerated carriers, especially for long-wavelength photons that penetrate deep into the absorber and reach the backside. As illustrated further below (**Fig. 5d**), this passivation increases the current density by 1.5 mA/cm$^2$ for the solar cell with an absorber thickness of (0.2+0.9) μm (SCS30). For the solar cells with thinner absorber of (0.10+0.75) μm (SP30 and the corresponding Mo-Re)), the enhancement of the long-wavelength EQE is much less pronounced. This reduced effect of the passivation can be attributed to non-absorption losses, which reduce the effect of passivation on the EQE.

We investigate In$_2$O$_3$ layers of different thicknesses, all with Cu annealing, and obtain similar passivation effects to the 30 nm In$_2$O$_3$, leading to nearly identical TRPL decay and EQE responses, as depicted in **Extended Data Fig. 3**. The robustness of HTL is encouraging, as it enables tolerance to thickness variations. Taking this advantage, precise control of In$_2$O$_3$ thickness within a few nanometer ranges becomes unnecessary, which holds significant implications for large-scale manufacturing by enhancing reproducibility. Furthermore, it will be demonstrated later that the high FF is only minimally impacted by In$_2$O$_3$ thickness, when Cu-annealing is provided, underscoring its potential for large-scale application.

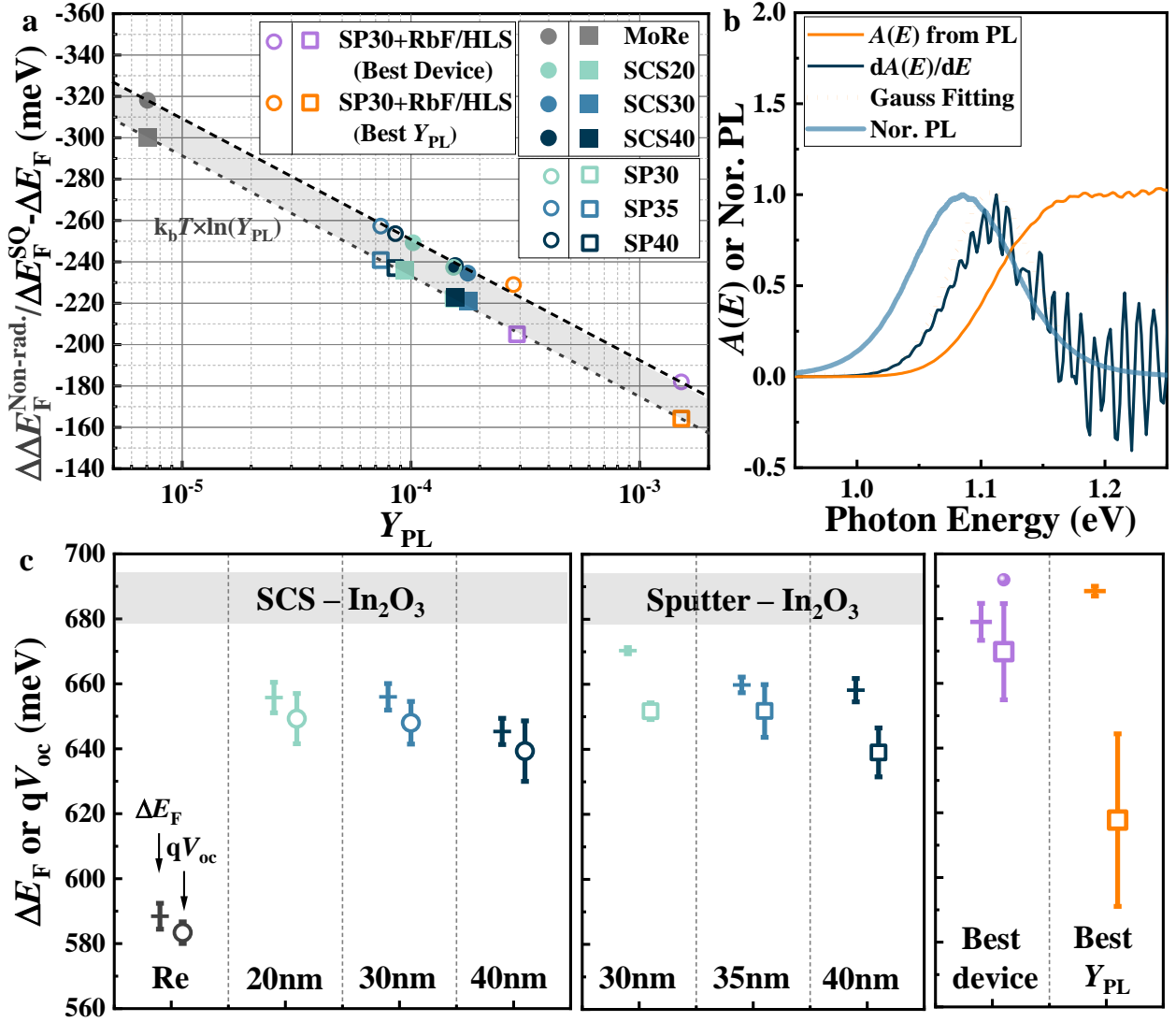

**Fig. 4| a,** Non-radiative quasi-Fermi level splitting ΔE_F loss [$\Delta\Delta E_F^{Non-rad.} = k_b T \times \ln Y_{PL}$] (squares) and independently measured quasi-Fermi level splitting ΔE_F deficits with respect to the Shockley-Queisser (SQ) limit ($\Delta E_F^{SQ} - \Delta E_F$) (circles) as a function of $Y_{PL}$. Solid symbols: solution processed HTLs and the Mo reference, open symbols: sputter processed HTLs. The non-radiative loss is also described by the grey line. The dark dashed line is a fit to the ΔE_F loss and is about 16 meV higher than the non-radiative loss due to additional radiative loss. Additional samples with RbF post-deposition treatment and heat-light soaking (HLS) are given. **b,** Absorptance spectrum A(E), derivative of the absorptance spectrum dA(E)/dE together with a Gaussian fit (normalised) and normalized PL spectrum of the sample with the best $Y_{PL}$. **c,** ΔE_F and qV_oc for different solar cells. Both types of HTL show similar improvement in ΔE_F and qV_oc

of ~70 meV in comparison to the reference solar cells, The somewhat lower q$V_{oc}$ compared to the $\Delta E_F$ is most likely due to the front surface recombination. The high $V_{oc}$ deficit of the sample with best $Y_{PL}$ is caused by the blocked diode current.

All samples are analysed by absolute PL. The effective passivation yields a notable enhancement in $Y_{PL}$ of more than an order of magnitude, consequently reducing the non-radiative $\Delta E_F$ loss by 60~80 meV relative to the reference sample. As depicted in **Fig. 4a** by the square symbols, the non-radiative $\Delta E_F$ losses are derived from $Y_{PL}$ according to [$k_b T \cdot \ln Y_{PL}$]. Both solution and sputter prepared HTLs exhibit considerably lower non-radiative $\Delta E_F$ loss than the reference sample on Mo, the non-radiative losses come down to 220~240 meV. No correlation between $In_2O_3$ thickness and non-radiative $\Delta E_F$ loss is evident. This absence of a trend is likely attributable to differences in the absorber or front surface passivation at the interface with the CdS buffer. We often observe differences of 10 meV in $\Delta E_F$ for different samples, prepared by supposedly identical conditions. Thus, we conclude that the effect of backside passivation tends to saturate when the $In_2O_3$ thickness is 20 nm or more, where the backside recombination is reduced to a level that has little impact on $\Delta E_F$ of our samples with lifetimes in a range of 100 to 200 ns. However, the level of backside recombination could become more critical if the absorber's bulk lifetime is enhanced through other techniques, such as Ag alloying or heavy alkali PDT. Therefore, it appears safer to use HTLs based on $In_2O_3$ layers, somewhat thicker than the minimum thickness of 20 nm. This is the reason why our samples with RbF PDT employ a 30 nm thickness of $In_2O_3$ – see below.

Besides the non-radiative loss from $Y_{PL}$, absolute PL allows the direct and independent determination of $\Delta E_F$ from a fit of the PL spectra to Planck's generalized law. The $\Delta E_F$ deficit is then defined as ($\Delta E_F^{SQ} - \Delta E_F$), where $\Delta E_F^{SQ}$ represents the quasi Fermi-Level splitting of the Shockley-Queisser (SQ) limit. The bandgap used to determine $\Delta E_F^{SQ}$ is taken as the inflection point of the onset of the absorptance spectrum [$A(E)$][25]. The absorptance spectrum is extracted from the PL spectrum, using Planck's generalised law, as explained in detail in the methods section. As an example, **Fig. 4b** displays $A(E)$ and $dA(E)/dE$ of the sample with the best $Y_{PL}$. Due to the noise in $dA(E)/dE$, a Gaussian fit is applied to the noisy curve, as shown (after normalisation) by the orange doted line. We also display the normalised PL spectrum. Its maximum is shifted to lower energies than the band gap by about 25 meV, which is expected based on the gradual absorption edge[26,55], The gradual absorptance increase can be attributed to a Ga gradient[56,57], which is not the case here, to tail states[56,58,59] and potential fluctuations[60-63]. The smoother the absorptance edge is the larger is the radiative $\Delta E_F$ loss as well as the shift between band gap and PL maximum[26,61]. The measured $\Delta E_F$ is reduced from the SQ limit by radiative and non-radiative losses. The complete loss is depicted in **Fig. 4a** by circles and by the dark dashed line. All the points lie almost on the line. The dark dashed line is 16 meV above the (grey) line of the non-radiative loss. This difference indicates a constant value of the radiative loss of 16 meV in addition to the non-radiative loss. The low value of the radiative loss is attributed to the absence of a gradient (**Fig. 1i**). The radiative loss here is much lower than the reported value of 36 meV for traditional CIGS with a Ga gradient[25]. Optimizing the absorber to be fully homogeneous in depth and closer to stoichiometry may further reduce the radiative $\Delta E_F$ loss by several meV. However, this improvement is relatively small when considering the non-radiative $\Delta E_F$ loss. Thus, the $\Delta E_F$ loss in CIGS is by far dominated by non-radiative loss, as was observed before[30,64].

Quasi-Fermi level splitting gives us the potential $V_{oc}$ of the finished solar cell, it is of course essential to measure $V_{oc}$ in actual solar cells. **Fig. 4c** demonstrates improvement in $V_{oc}$ for HTL-passivated solar cell, very similar to the $\Delta E_F$ improvement. Both solution and sputter deposited $In_2O_3$ result in improved $V_{oc}$ ranging between 640 to 660 mV, about 70 mV higher than the $V_{oc}$ of the reference sample (~580 mV). This reaffirms that sputtered $In_2O_3$ is a viable alternative to solution prepared $In_2O_3$, thus paving the way for potential large-scale manufacturing applications. While the $qV_{oc}$ of solar cells with sputter $In_2O_3$ is marginally lower than their corresponding $\Delta E_F$, this discrepancy is likely attributed to front surface recombination, which tends to lower the quasi-Fermi level of electrons toward the front contact[44]. From numerous Cu(In,Ga)Se$_2$ solar cells, prepared at different labs and measured by our lab, we find in reasonably good devices difference between $V_{OC}$ and $\Delta E_F$ that vary from 5 to 20 meV[44,65], potentially stemming from systemic deviations in device preparation. However, this variation does not affect the conclusion that HTL passivation substantially enhances $V_{oc}$ compared to reference solar cells.

Achieving high-efficiency solar cells relies on achieving both effective passivation (yielding high $V_{oc}$) and efficient hole transport (resulting in high FF). The *J-V* curves of the reference solar cell and those passivated by Cu-annealed HTL, employing either 30 nm solution or sputter prepared $In_2O_3$, are presented in **Fig. 5a**. The diode character of the *J-V* chracteristics indicates comparable FF to the reference cell, coupled with a noticeable increase in $V_{oc}$. $V_{oc}$ and FF distributions are summarized in **Fig. 5b and c**, respectively. The passivated samples exhibit a $V_{oc}$ enhancement of 70 mV and good FF ranging from 70% to 76%. As a result, the efficiency is elevated from ~14% to over 16% for solar cells with solution prepared $In_2O_3$. Solar cells with sputtered $In_2O_3$ demonstrate a slightly lower efficiency of ~15.5% due a lower current, which might be attributed to the thinner absorber (0.1+0.75 μm), leading to $J_{sc}$ losses due to non-absorption, as illustrated in **Fig. 5d** and **Fig. 3b**. These outcomes underscore the effectiveness of Cu annealing in improving hole transport properties through the final GaOx film while maintaining adequate passivation, thereby paving the way for highly efficiency homogeneous CIGS solar cells with reduced thickness.

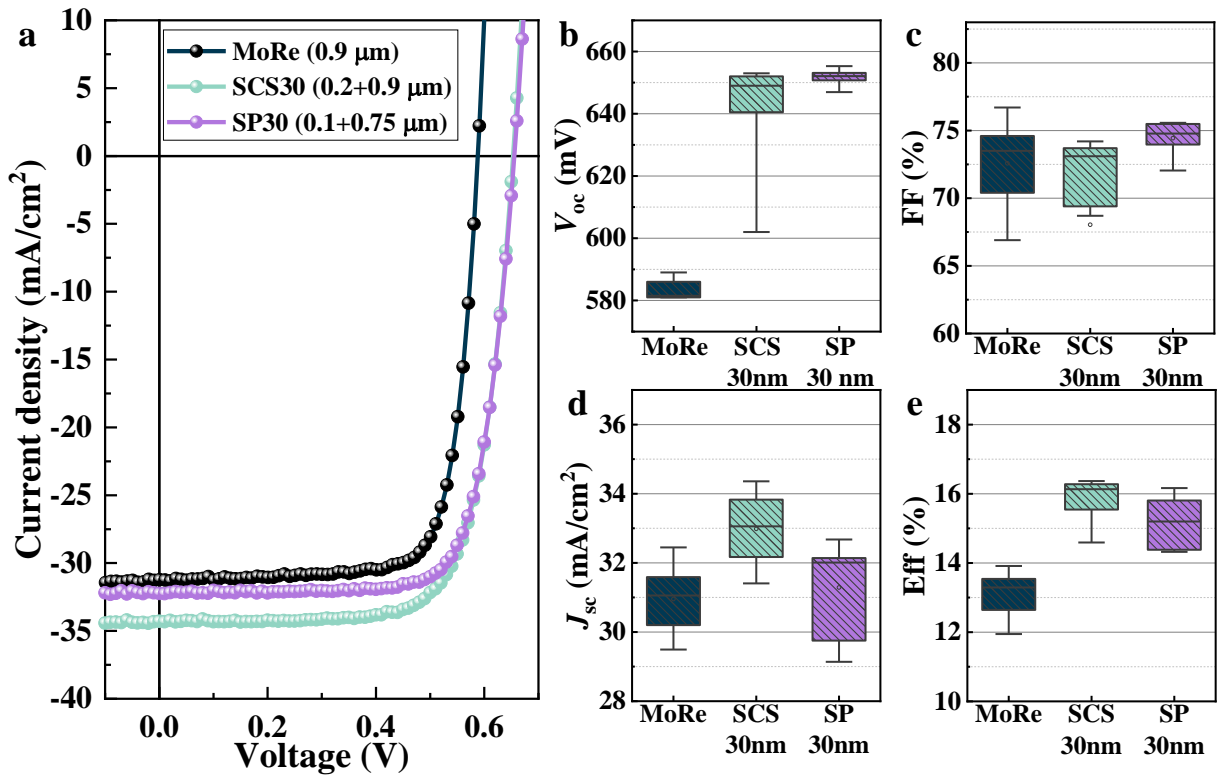

**Fig. 5| a,** *J-V* characteristics under illumination of the reference solar cell and solar cells passivated by Cu-annealed HTL with solution or sputter prepared 30 nm $In_2O_3$. Cu annealing is effective to improve hole transport property of HTL, leading to higher FF. Here we show the best solar cells from each experiment. The statistical results of solar cell parameters: **b,** open circuit voltage; **c,** fill factor; **d,** short circuit current and **e,** active area efficiency.

Varying thicknesses of $In_2O_3$ lead all to solar cells with a notable and comparable improvement in $V_{oc}$, as depicted in **Fig. 4c**. The majority of these solar cells demonstrate robust FFs ranging from 70% to 76%, as illustrated in **Extended Date Fig. 4** and **Extended Date Fig. 5**. This indicates that Cu annealing is an effective way to make the GaOx layer in the HTL conductive for holes. However, further investigation into the impact of annealing conditions—such as Cu quantity, vacuum level, annealing duration, and substrate temperature—on HTL properties may further improve the solar cells.

**Efficiency improvement introduced by adding RbF PDT and heat light soaking**

The Cu-annealed HTL demonstrates effective passivation and hole transport properties, thereby shifting the dominant efficiency limitation from backside recombination to bulk recombination, as previously discussed[30]. To enhance solar cell efficiency further, we have introduced RbF PDT and heat light soaking (HLS)[32,38,66,67]. Initial outcomes reveal promising enhancements in $\Delta E_F$. As depicted in **Fig. 4a**, by adding RbF PDT and heat light soaking, we achieved a $Y_{PL}$ of 0.15% with a corresponding non-radiative $\Delta E_F$ deficit of approximately 165 meV, which is notably 60 meV lower than samples without RbF PDT and heat light soaking, suggesting a pathway to improve solar cell performance. However, the resultant solar cells exhibit S-shaped $J$-$V$ curves with reduced FF, $V_{oc}$ and $J_{sc}$, as outlined in **Extended Date Fig. 6**. This phenomenon may stem from excessive RbInSe$_2$ thickness due to an unoptimized RbF PDT, impeding electron transport[68]. Additionally, we cannot exclude the possibility that overdosing Rb and diffusion to the backside might change the HTL properties, consequently hindering hole transportation.

Further optimisation of the RbF yielded the best device in this study. As illustrated in **Fig. 4a**, our top-performing solar cell exhibits suboptimal non-radiative $\Delta E_F$ loss of 205 meV, approximately 40 meV higher than the optimal case. However, the solar cell demonstrates a good FF of ~76% (**Fig. 6a**), resulting in an in-house measured (active area) efficiency of 18.2% with ARC. We had this cell certified at ISE CalLab and obtained a certified full-area efficiency of 16.2% (certified $J$-$V$ curve is shown in **Extended Date Fig. 7**). The difference is mainly due to our unoptimized grids which causes around 10% of shadowing loss to the $J_{sc}$. This efficiency ranks among the highest reported for sub-micron CIGS solar cells, as summarized in **Fig. 6b**. With further optimization, achieving efficiencies close to 20% for sub-micron CIGS (0.1+0.75 μm) and beyond 21% for slightly thicker CIGS (e.g., 1.0 μm) seems very possible.

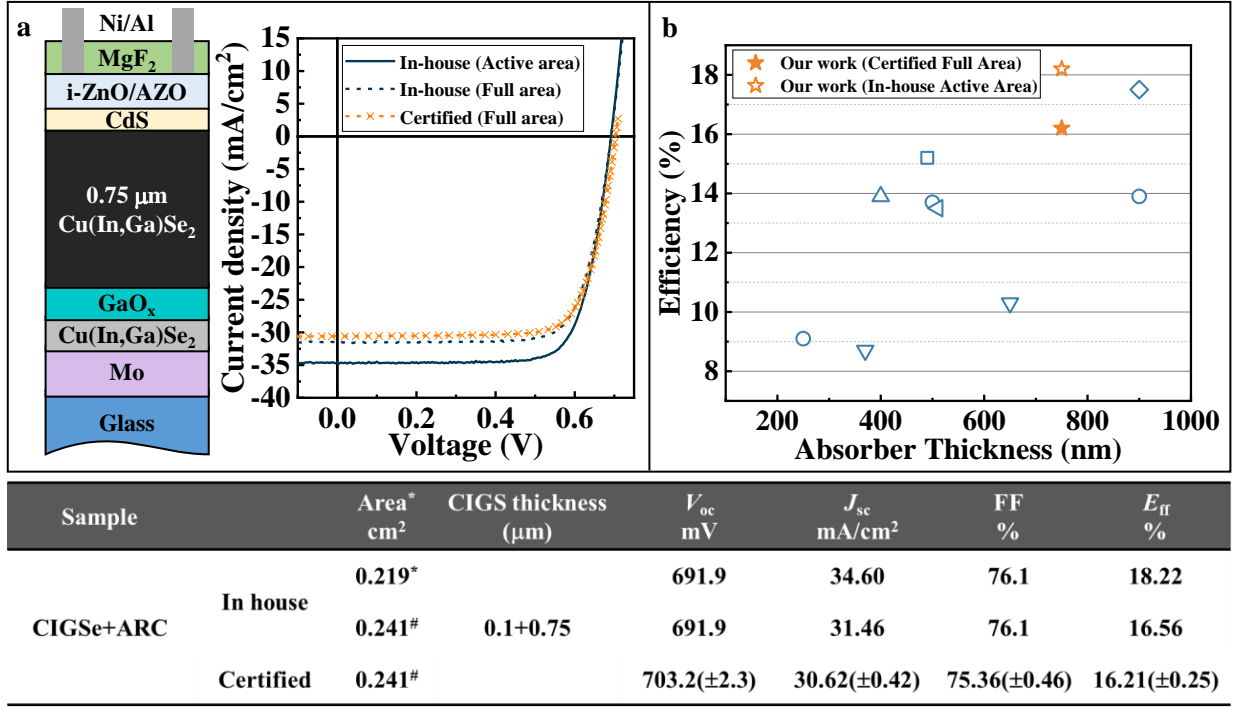

| Sample | | Area* cm$^2$ | CIGS thickness (μm) | $V_{oc}$ mV | $J_{sc}$ mA/cm$^2$ | FF % | $E_{ff}$ % |
|---|---|---|---|---|---|---|---|
| CIGSe+ARC | In house | 0.219* | 0.1+0.75 | 691.9 | 34.60 | 76.1 | 18.22 |
| | | 0.241# | | 691.9 | 31.46 | 76.1 | 16.56 |
| | Certified | 0.241# | | 703.2(±2.3) | 30.62(±0.42) | 75.36(±0.46) | 16.21(±0.25) |

*: Active area: Full area – grids; #: Full area (including grids)

**Fig. 6| a,** The device structure plus the in-house and certified *J-V* characteristics of the record sub-micron solar cell. **b,** The efficiency of this record sub-micron CIGS cell compared to the efficiency of the best literature reported sub-micron CIGS solar cells. Circle[69], square[15], down triangle[70], up triangle[48], left triangle[20], diamond[19].

**Conclusion**

A novel HTL structure has been implemented in sub-micron CIGS solar cells, initially comprising CGS/In$_2$O$_3$ layers, which transforms into CIGS/GaO$_x$ during absorber growth. This bi-layer provides effective backside passivation without hampering the hole transport. The backside passivation yields a notable increase in $V_{oc}$ by 60-80 mV. In the absence of Cu annealing of the HTL, solar cells exhibit markedly low FFs, attributed to inadequate hole transport in the HTL. However, Cu annealing significantly enhances FFs to 77%, indicating a crucial enhancement in hole transport through the HTL. Moreover, the performance of sputtered In$_2$O$_3$ matches the one of solution prepared In$_2$O$_3$, facilitating potential large-scale manufacturing applications. With In$_2$O$_3$ thickness ranging from 20 to 40 nm, the passivation and hole transport of Cu-annealed HTL are hardly impacted by In$_2$O$_3$ thickness variation. This flexibility eliminates the need for precise control within a few nanometers range, enhancing feasibility and adaptability for large-scale manufacturing. Furthermore, with additional RbF PDT and heat light soaking, we achieved a notable $Y_{PL}$ of ~0.15% (non-radiative $\Delta E_F$ loss = ~165 meV) and established a best active area efficiency of 18.2% for sub-micron CIGS solar cells with an absorber thickness of 0.75 μm.

**Methods**

*Sample preparation*

**Mo:** 500 nm molybdenum is prepared by sputtering.

**CuGaSe$_2$:** 100~200 nm CuGaSe$_2$ is deposited by co-evaporation with a substrate setting temperature of 356 °C.

**In$_2$O$_3$:** The In$_2$O$_3$ is prepared by solution combustion synthesis or sputtering.

Solution prepared: The precursor solution is formulated by dissolving 1203.2 mg of In(NO$_3$)$_3$·xH$_2$O (99.99%, Sigma-Aldrich) in 20 mL of 2-methoxyethanol (2-MOE, 99.8%, Sigma-Aldrich), resulting in a 0.2 M concentration. Subsequently, 800 μL of acetylacetone (C$_5$H$_8$O$_2$, 99%, Sigma-Aldrich) is introduced as a fuel, followed by the addition of 360 μL of 14.5 M NH$_3$ (aqueous, 99%, Sigma-Aldrich) to adjust the pH and facilitate the formation of In(acac)$_x$ (acac = C$_5$H$_7$O$_2$) complexes of In ions. The solutions are then agitated until they achieve clarity. Using this clear solution, In$_2$O$_3$ films are produced by spin-coating onto substrates at 3000 rpm for 60 s, followed by hot-plate heating at 130 °C for 1 min. This spin-coating and drying process is repeated to attain film thicknesses of 20 nm, 30 nm, and 40 nm. Finally, the films are crystallized by placing the samples on a hot-plate in air at 300 °C for 3 min.

For sputtered passivated samples, the indium oxide (In$_2$O$_3$) layers are deposited using the RF sputtering technique. Various thicknesses of In$_2$O$_3$ are achieved by adjusting the deposition energies: 25 nm at 39 kJ, 30 nm at 46 kJ, 35 nm at 54 kJ, and 40 nm at 62 kJ.

**Cu annealing of the HTL:** Around 40 nm of Cu is deposited onto the surface of In$_2$O$_3$ at a substrate setting temperature of 200 °C. The temperature is then quickly increased to 500 °C at a rate of 50 °C per minute, after which a 20-minute annealing process is conducted under vacuum conditions of approximately $5\times10^{-9}$ Torr inside the MBE chamber that we use for absorber deposition. Next, the substrate temperature is gradually decreased to 356 °C at a rate of 20 °C per minute, after which the standard 3-stage process for CIGS preparation commences. For the high temperature process discussed below, the substrate temperature remains constant at 500 °C for the 1st stage, with no temperature reduction occurring.

**Cu(In,Ga)Se$_2$:** The absorber undergoes a standard 3-stage preparation process. For the best solar cell with efficiency of 18.2%, initially, the In-Ga-Se is supplied at a substrate temperature of 500 °C in the first stage. For the other samples, this temperature is 356 °C. Subsequently, during the second stage the substrate setting temperature is heated up to 580 °C at a rate of 20 °C per minute with co-evaporation of Cu and Se. Upon reaching a Cu-rich state (Cu/[In+Ga] ≈ 1.2, estimated based on the process times, e.g., from the 1st stoichiometric point. The 1st stoichiometric point is where the output power of substrate heater starts increasing), the Cu shutter closes, and the film undergoes a 5-minute annealing process in Se atmosphere. In the third stage, maintaining the substrate temperature at 580 °C, In, Ga are reintroduced under Se pressure to achieve a final absorber state slightly Cu-deficient (Cu/[In+Ga] = 0.92~0.95, determined using EDS) with a thickness of approximately 0.75 μm to 0.9 μm. Notably, the pyrometer readings for substrate temperature typically register lower values than the setting temperature. This discrepancy tends to increase with higher setting temperatures; for instance, at 365 °C, both readings align closer, while at 580 °C, the pyrometer indicates a temperature 50~60 °C lower than the setting temperature. During the absorber growth processes, the Se flux is around $4.5 \times 10^{-6}$ Torr which is determined by an ion gauge facing to the Se source before and after the absorber growth.

**RbF PDT:** The RbF PDT is conducted after the absorber deposition without interrupting the vacuum. Following the absorber deposition, the substrate setting temperature is lowered to 280 °C. From the pyrometer, the reading temperature is around 313±5 °C. A subsequent 10-minute RbF PDT is performed by supplying RbF for 10 minutes under a Se atmosphere. The Se flux is around $2.5 \times 10^{-6}$ Torr, the RbF source temperature is 450 °C.

**CdS:** The CdS layer is fabricated using chemical bath deposition. Prior to the CdS application, all samples undergo a 30-second chemical etching process using a 5% aqueous KCN solution, aimed at eliminating potential residual oxides. For the samples with RbF PDT, they are additionally etched by the 1.5 M NH$_4$OH for 3 minutes after the KCN etching. The chemical bath process involves deposition for 6-7 minutes at 67°C, utilizing a solution composed of 2 mM CdSO$_4$, 50 mM thiourea, and 1.5 M NH$_4$OH. Based on standard growth rates, the estimated thickness of the CdS layer ranges between 40 to 50 nm. This CdS layer plays a crucial role in passivating the front surface, effectively preventing surface degradation.

**TCO and grids:** To complete the device assembly, the i-ZnO/ZnO:Al layers are sequentially RF-sputtered atop the CdS layer, followed by the deposition of Ni/Al grids via e-beam evaporation. The transparent conducting oxide (TCO) deposition process was conducted using a commercial semi-automated sputtering deposition system. Two magnetron guns, outfitted with ceramic 3-inch diameter by 0.125-inch thick ZnO and ZnO:Al (2 wt% Al) targets, were powered by RF generators and operated within a non-reactive Ar atmosphere. The deposition of the non-conductive ZnO (i-ZnO) and conductive ZnO:Al (AZO) films involved applying sputtering powers of 125 and 140 W to the respective targets, while maintaining a pressure of 1 mTorr. The resulting thicknesses of the i-ZnO and AZO films are around 80 and 380 nm, respectively. Ni-Al grids are deposited by e-beam evaporation.

**MgF anti-reflection coating (ARC):** 90 nm MgF ARC layer is evaporated by an e-beam evaporator onto the top of the TCO layer and the grids to reduce the reflection losses.

**Heat-light-soaking:** The heat light soaking is done with a substrate setting temperature of 80 °C in a $N_2$ atmosphere and under equivalent illumination intensity of 0.5 Sun for 3 hours. The samples are sealed in a transparent plastic bag that is purified and refilled with 99% pure dry $N_2$. Then samples are placed on a hot plate with setting temperature of 80 °C under a LED light source (EACLL GU10-1050 Lumen and 6000 K bulbs) with equivalent illumination intensity of 0.5 Sun, which means that the photons flux with energy above 1.12 eV is roughly half of that given by AM1.5G spectrum.

*Characterization*

**Time-resolved photoluminescence (TRPL):** This technology relies on time-correlated single photon counting (TCSPC) to analyze luminescence decays in the time domain. Measurements are conducted using a 640 nm pulsed diode laser. For samples exhibiting short or long lifetimes, the laser repetition rate is set to 5 or 1 MHz respectively. The typical average power of the laser at a repetition rate of 2 MHz is 0.25 mW, with a laser diameter of approximately 0.8 mm. To mitigate the pile-up effect and prevent loss of long-lifetime photons, the ratio between total count rate and repetition rate is maintained below 2% by adjusting the neutral density (ND) filters to modulate laser intensity if necessary. A 2-exponential decay function is employed to fit the PL decay curve:

$$I = I_0 + A_1 \exp\left(\frac{-t}{\tau_1}\right) + A_2 \exp\left(\frac{-t}{\tau_2}\right) \tag{1}$$

In which, $I$ represents the intensity of the PL counts, while $I_0$ is the fitted background counts. The parameters $\tau_1$ and $\tau_2$ correspond to the fitted lifetimes for the fast and slow decays, respectively. In our investigation, we specifically focus on $\tau_2$, which is significantly impacted by backside recombination[41].

**Absolute PL:** The absolute PL is measured using a home-built setup. All samples are excited by a diode laser with a wavelength of 660 nm and evaluated in ambient air at room temperature. The laser has an approximate diameter of 2.6 mm. Initial collection of photoluminescence involves two parabolic mirrors, redirecting the light to a monochromator via a 550 μm optical fiber. An InGaAs array detector captures the emitted light. The obtained PL spectra undergo spectral correction through calibration by a halogen lamp with know spectrum. Quantification of both excitation and corrected radiation flux is achieved using a power meter, enabling calculations of $\Delta E_F$ across specific illumination intensities ranging from 0.01 sun, or even lower, up to several sun equivalents, contingent upon the absorbers' quality and $E_g$. Here, one sun intensity signifies that the photon flux matches the AM1.5 spectrum above the absorber's bandgap $E_g$. Applying Planck's generalized law[71]:

$$\phi_{PL}(E) \approx A(E)\phi_{bb}(E)\exp\left(\frac{\Delta E_F}{k_b T}\right) \tag{2}$$

In which, $\phi_{PL}(E)$ is the measured PL spectrum, $\phi_{bb}(E)$ is the black body radiation, $A(E)$ is the absorptance and $k_b T$ is the thermal energy. With the temperature fixed at the independently measured 295 K[26,43], $\Delta E_F$ is computed by fitting the high-energy wing of the PL spectra, assuming absorptance ($A(E)$) equals 1 in this energy range. However, it is important to note that this assumption leads to a slight underestimation of $\Delta E_F$. Recent discussions have highlighted the impact of $A(E) <1$ for high-energy photons[26]. With fitted $\Delta E_F$, the $A(E)$ of the sample can be re-calculated from **Equation 2**. Because the 1st and 2nd diodes in our InGaAs array detector have different dark counts, which leads to non-smooth curves in a jagged shape, and becoming serious when the PL signal is low. To have a smoother $A(E)$, thus having a better $dA(E)/dE$, the $A(E)$ shown in **Fig. 4b** only contains the data that are collected from the second diode. It means that we have removed the half of the pixels to minimize the influence of high dark counts on low PL signal.

$$A(E) = \frac{\phi_{PL}(E)}{\phi_{bb}(E)} \exp\left(-\frac{\Delta E_F}{k_b T}\right) \tag{3}$$

The PL quantum yields ($Y_{PL}$) is determined by the ratio between the absolute incident photon flux and PL flux. This approach also neglects reflection of laser photons and slightly underestimates $Y_{PL}$. The incident photon flux is determined by first measuring the actual power of the laser spot using a power meter. Subsequently, considering that the photon intensity conforms to a Gaussian distribution within its diameter, which is determined by a CCD camera. The PL flux is then derived by integrating the absolute PL spectra across its emitted photon energy range. With knowing $Y_{PL}$, the non-radiative loss in $\Delta E_F$ ($\Delta\Delta E_F$) can be determined by[26,42,43]:

$$\Delta\Delta E_F = k_b T \times \ln(Y_{PL}) \tag{4}$$

**Illumination current density-voltage (*J-V*):** The measurements are conducted at 25°C using a 4-probe configuration. A class AAA solar simulator provided a simulated AM1.5G spectrum, calibrated using a Si reference cell. During the assessment, a forward scanning voltage ranging from -0.3 to 0.8 V is applied incrementally at intervals of 0.01 V with a waiting time of 0.25 s and scanning speed of 1 V/s. The certification measurements are done by Fraunhofer ISE CalLab PV Cells and certification can be found in **Extended Date Fig. 7**.

**External quantum efficiency (EQE):** The EQE spectra are acquired using a home-built setup featuring a grating monochromator configuration under chopped illumination from halogen and xenon lamps. A lock-in amplifier facilitated the measurement of the solar cell's photocurrent. Calibration reference spectra are obtained using a calibrated Si detector covering the range of 300 to 1100 nm and a calibrated InGaAs detector spanning 1100 to 1400 nm. The measured solar cells are connected using 4 pins and measured in 2-probe configuration.

**Scanning electron microscope (SEM):** Scanning electron microscope is used to analyze the cross-sectional microstructures of the films.

**Transmission electron microscopy (TEM):** In this study, cross-section Transmission Electron Microscopy (TEM) samples are meticulously prepared using a Focused Ion Beam (FIB) system, specifically the FEI Helios Nanolab 650. The TEM analysis is conducted using a JEOL F200-Cold

FEG instrument. Elemental mapping and profiling are accomplished via X-ray Energy Dispersive Spectroscopy (EDS) in Scanning Transmission Electron Microscopy (STEM) mode.

**Secondary ion mass spectrometer (SIMS):** Measurements are conducted using a CAMECA SC-ultra instrument (Ametek). A 1 keV focused $Cs^+$ ion beam (16 nA) is utilized to sputter across a sample surface area measuring 500 µm × 500 µm. Only positive ions originating from the central region, with a diameter of 30 µm, are detected as $MCs^+$ or $MCs_2^+$, where M represents the ions of interest, such as Cu, In, Ga, Se, O, and Mo. Data are plotted against sputtering time that is related to depth in the stack.

**Data Availability Statement**

The data that support the findings of this study can be obtained from the corresponding author.

**Acknowledgements**

This work was supported by the Luxembourg National Research Fund (FNR) through the PACE project under the grant number of PRIDE17/12246511/PACE and by the European Union in the framework of the HiBITS project. The authors thank Brahime El Adib for his technical assistance in SIMS characterization. ChatGPT, a language model developed by OpenAI in San Francisco, CA, USA, provided assistance in English language editing. The whole text has been carefully modified and verified by the authors. For the purpose of open access, the author has applied a Creative Commons Attribution 4.0 International (CC BY 4.0) license to any Author Accepted Manuscript version arising from this submission.

**Author contributions**

T.W conceived the idea and designed the experiments. L.S prepared the oxides by solution combustion synthesis and wrote the part of paper about oxides preparation. T.W wrote the rest part of the paper. S.B. prepared the sputtered $InO_x$ layers. T.W and S.B prepared the $Cu(In,Ga)Se_2$ absorbers. T.W measured the samples by PL, conducted the *J-V* and EQE measurements. M.M made the most parts of the devices including glass cleaning, preparation of Mo, CdS, TCO and grids. M.M also conducted the SEM measurements. N.V and A.P conducted the TEM and SIMS measurements, results discussion and analysis, A.P also wrote the part about the TEM measurements. E.D, S.G, and S.S were involved in results discussion and analysis, contributing to the improvement of experiments. S.S defined the project, supervised this work and contributed to the writing of the paper. All the authors contributed to the revise of the paper.

**Competing interests**

The authors declare that they have competing interests related to the patent associated with this work. The patent in question is LU504697 (Application No.) titled "Thin film solar cell and corresponding production method". The inventor Taowen Wang is an employee of University of Luxembourg. The inventor Longfei Song is an employee of Luxembourg Institute of Science and Technology (LIST). Both University of Luxembourg and LIST hold the patent.

**Extended Data:**

Hole selective transport structure enhances the efficiency of
sub-micron Cu(In,Ga)Se$_2$ solar cell to 18.2%

Taowen Wang[1*], Longfei Song[2], Saeed Bayat[1], Michele Melchiorre[1], Nathalie Valle[2], Adrian-Marie Philippe[2], Emmanuel Defay[2], Sebastjan Glinsek[2], and Susanne Siebentritt[1*]

1. Laboratory for Photovoltaics (LPV), Department of Physics and Materials Science, University of Luxembourg, 41 rue du Brill, L-4422, Belvaux, Luxembourg.
2. Luxembourg Institute of Science and Technology (LIST), 41 rue du Brill, L-4422, Belvaux, Luxembourg

*Corresponding author: wangtaowenscu@hotmail.com , susanne.siebentritt@uni.lu

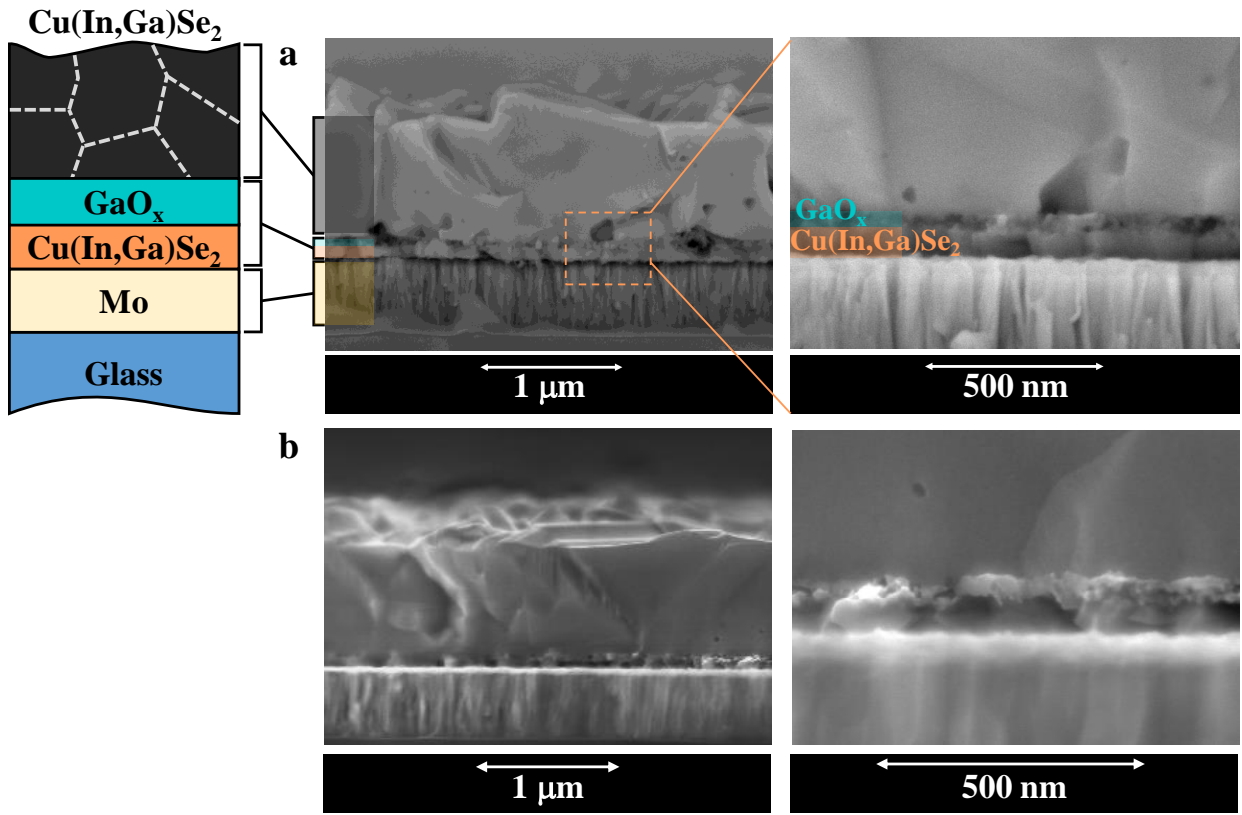

**Extended Data Fig. 1|** The sketch of the solar cell structure after the CIGS absorber deposition and its corresponding SEM cross-section images. The SEM cross-section image for samples: a, with solution deposited $In_2O_3$; b, with sputtered $In_2O_3$, which shows clearly that 2 individual layers stay at backside between Mo and CIGS, which indicates the thermal stability of the HTL. Before the absorber deposition, the stack of HTL is $CuGaSe_2/In_2O_3$, which changes to $CIGS/GaO_x$ after the absorber deposition as discussed in the main context.

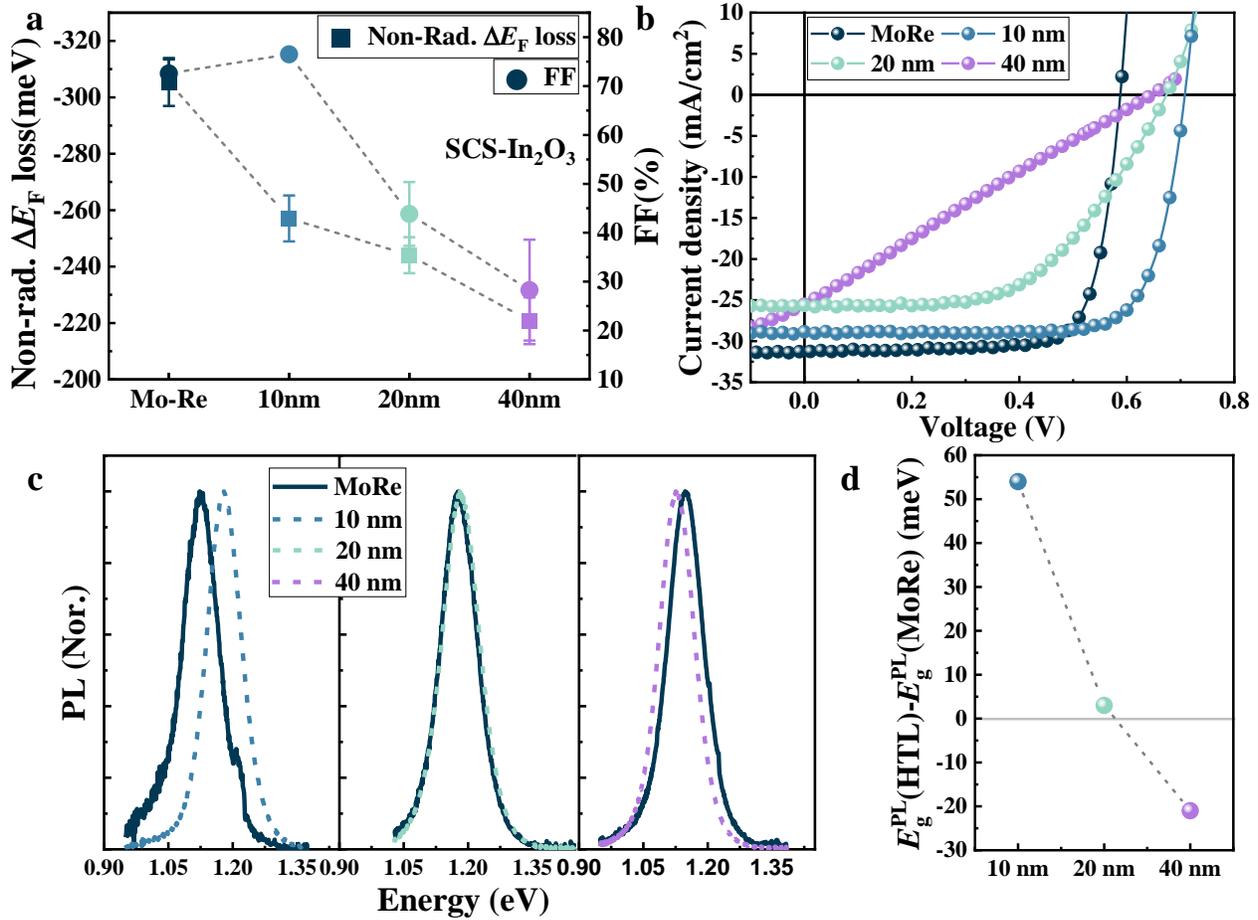

**Extended Data Fig. 2| The thickness optimization of In$_2$O$_3$. a,** Non-radiative $\Delta E_F$ loss and FFs of solar cells with and without HTL made from solution deposited In$_2$O$_3$ with thicknesses of 10 nm, 20 nm and 40 nm. Non-radiative $\Delta E_F$ loss are determined by k$_b$T*ln($Y_{PL}$). Due to the unpassivated back surface, the reference solar cell without HTL exhibits the highest non-radiative loss of $\Delta E_F$ of above 300 meV. Thicker In$_2$O$_3$ layers show better passivation, leading to lower non-radiative $\Delta E_F$ loss. However, the thicker In$_2$O$_3$ also results in dramatic reduction in FFs. **b,** Illuminated *J-V* curves of reference sample and samples with varying In$_2$O$_3$ thickness. It is clear that the thicker In$_2$O$_3$ results in lower FFs compared to the reference sample. Samples with 10 nm In$_2$O$_3$ exhibit similar FF to the reference sample, while those with 40 nm In$_2$O$_3$ show the lowest FF due to extensive hole transport blocking by the thick oxide layer. Moderate FF is observed in samples with 20 nm In$_2$O$_3$, suggesting partial hole transport blocking. Solar cells with thick oxide layers exhibit lower $J_{sc}$ due to current blocking. Addtionally, due to the current blocking, these solar cells show much lower q$V_{oc}$ compared to its $\Delta E_F$. **c,** Normalized PL flux of

samples passivated by the HTL with different thicknesses of $In_2O_3$. Comparing to their reference samples, the sample with 10 nm $In_2O_3$ shows a blue shift in the PL peak, which is most likely due to Ga diffusion from the $CuGaSe_2$ part of the HTL, which is blocked by thicker oxide layers. This can also properly explain its lower passivation compared to the thicker $In_2O_3$. **d,** The bandgap shift of HTL passivated samples compared to the reference sample.

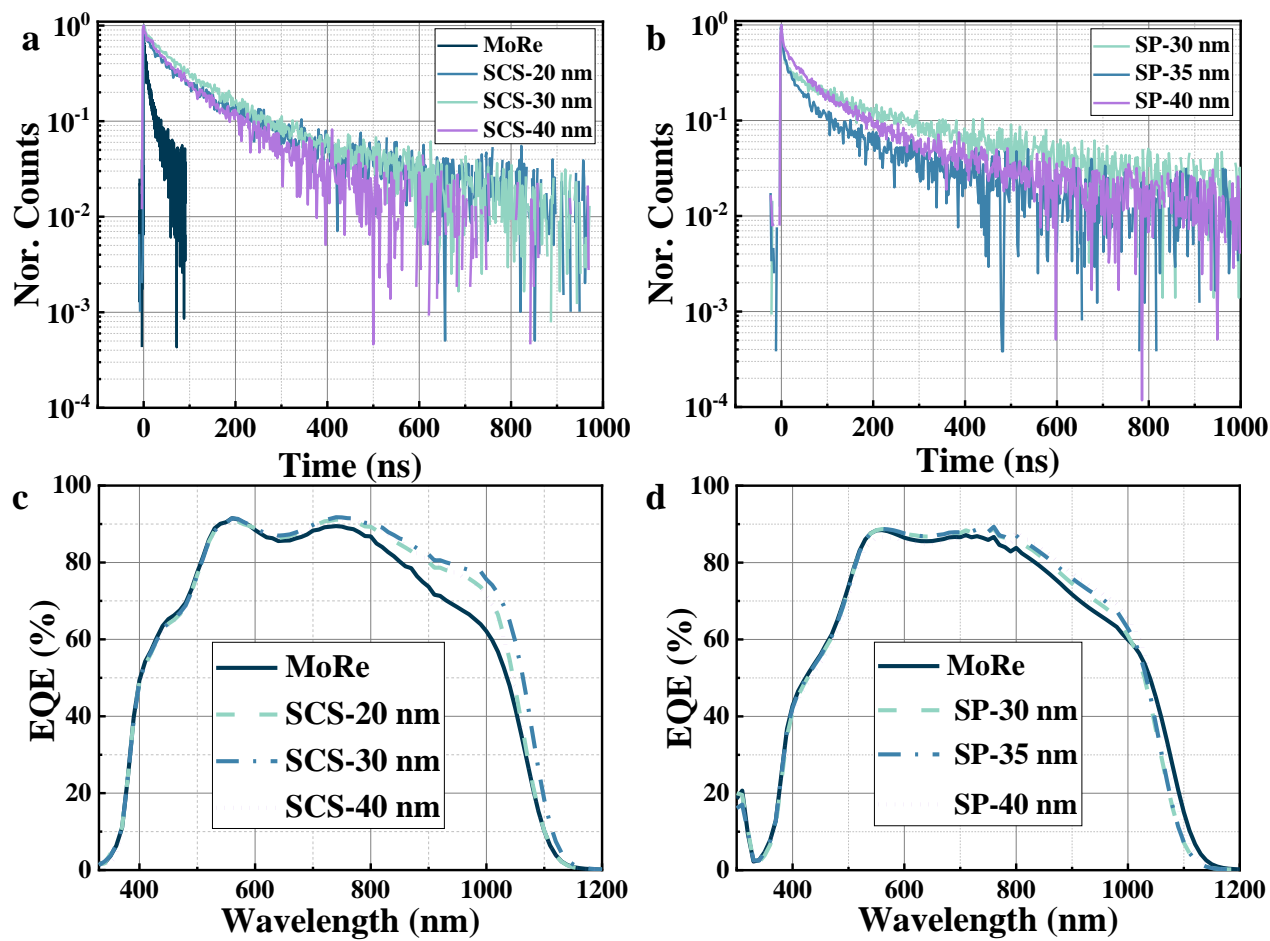

**Extended Data Fig. 3|** The TRPL decay (a and b) and EQE spectra (c and d) of samples passivated by HTL. **a, c,** HTL from solution deposited $In_2O_3$ with thicknesses of 20 nm, 30 nm, and 40 nm, **b, d,** HTL from sputtered $In_2O_3$ with thicknesses of 30 nm, 35 nm, and 40 nm. In all cases, similar slow decays of TRPL and enhanced collection of long-wavelength photons are observed, indicating equivalent passivation of backside recombination. However, solar cells with sputtered $In_2O_3$ exhibit less improvement in collecting long-wavelength photons, because they are based on thinner absorbers and show higher non-absorption losses.

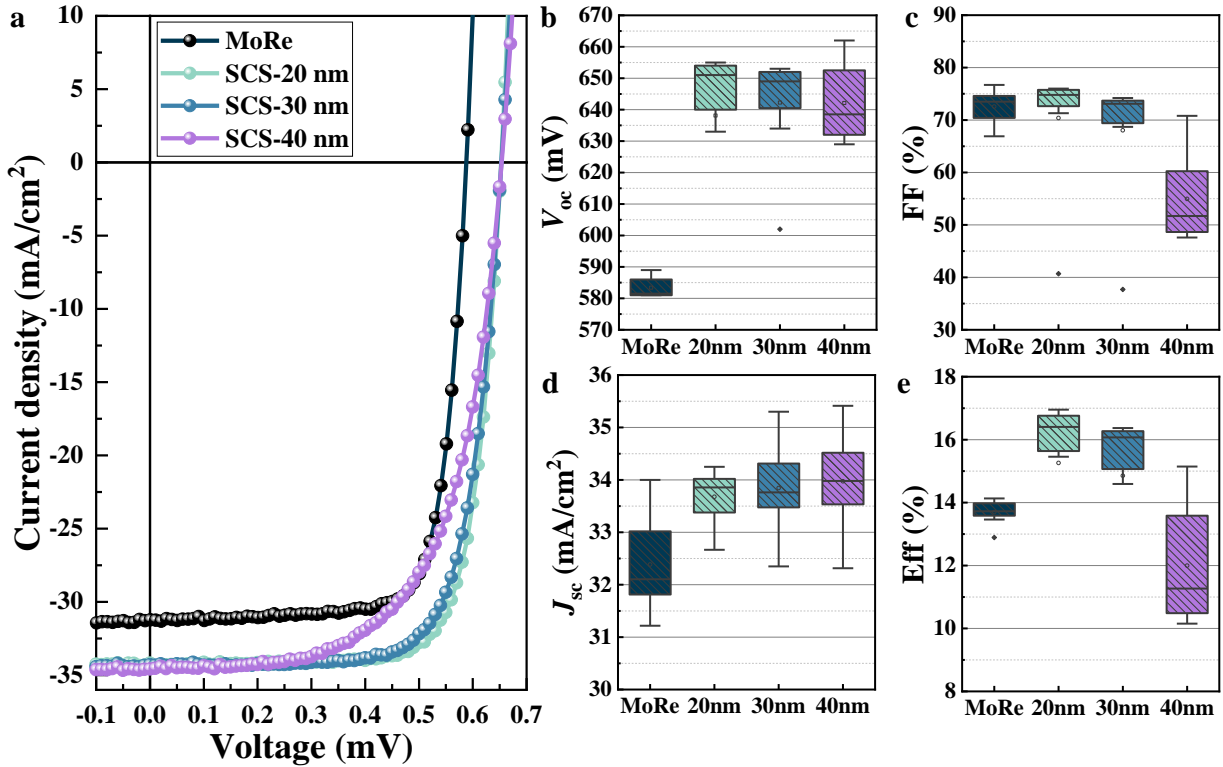

**Extended Date Fig. 4| a,** The *J-V* curves depict the performance of both the reference solar cell and the HTL passivated solar cells. The $In_2O_3$ layer of the HTL is fabricated using solution combustion synthesis, with thicknesses of 20 nm, 30 nm, and 40 nm. The *J-V* parameters of the solar cells are presented as follows: **b,** $V_{oc}$; **c,** FF; **d,** $J_{sc}$, and **e,** Eff.

The solar cells with 40 nm solution prepared $In_2O_3$ exhibit lower FFs of ~55% (**Extended Date Fig. 4**). Interestingly, solar cells with equivalent thicknesses of sputtered $In_2O_3$ do not display any FF issues. Furthermore, our prior research confirms that even 50 nm SCS-$In_2O_3$ can yield a commendable FF exceeding 71% for CIS solar cells[30]. Hence, we attribute this lower FF to the fact that the Cu-annealing process is not optimised, yet, particularly for the thicker solution prepared $In_2O_3$.

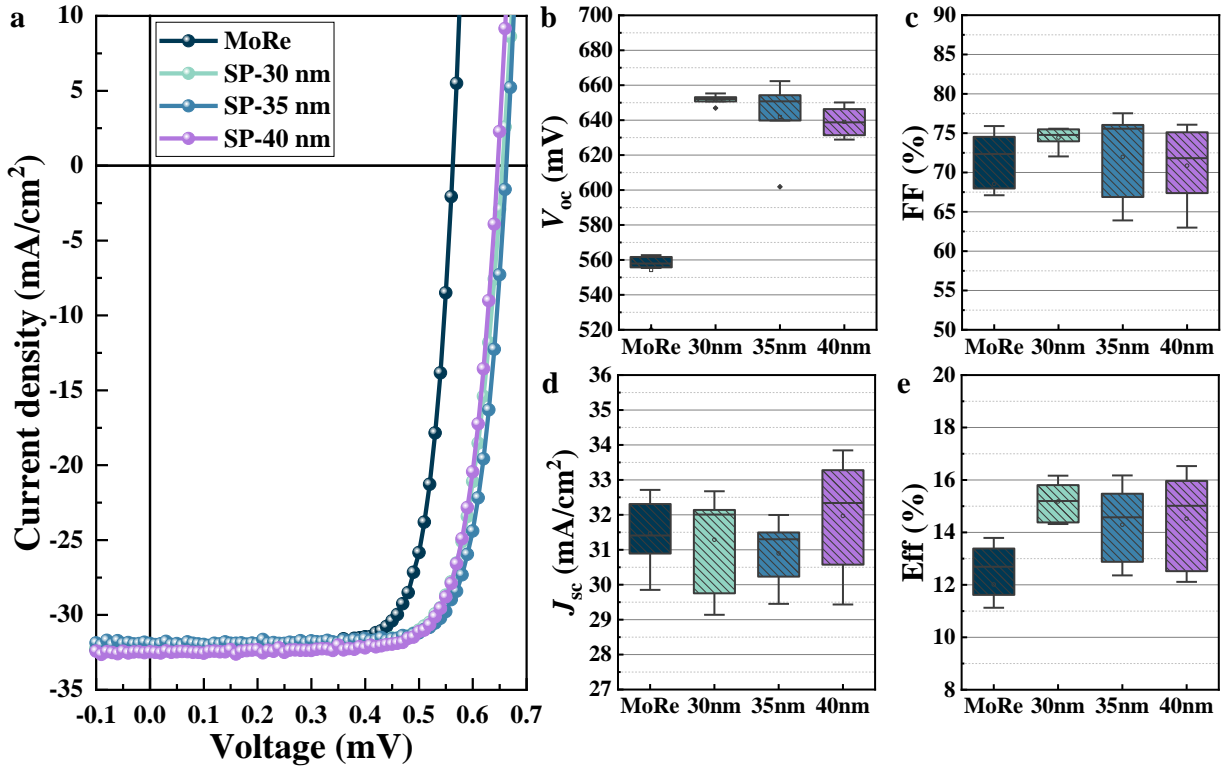

**Extended Date Fig. 5| a,** The *J-V* curves depict the performance of both the reference solar cell and the HTL passivated solar cells. The $In_2O_3$ layer of the HTL is fabricated using sputter, with thicknesses of 30 nm, 35 nm, and 40 nm. The *J-V* parameters of the solar cells are presented as follows: **b,** $V_{oc}$; **c,** FF; **d,** $J_{sc}$, and **e,** Eff.

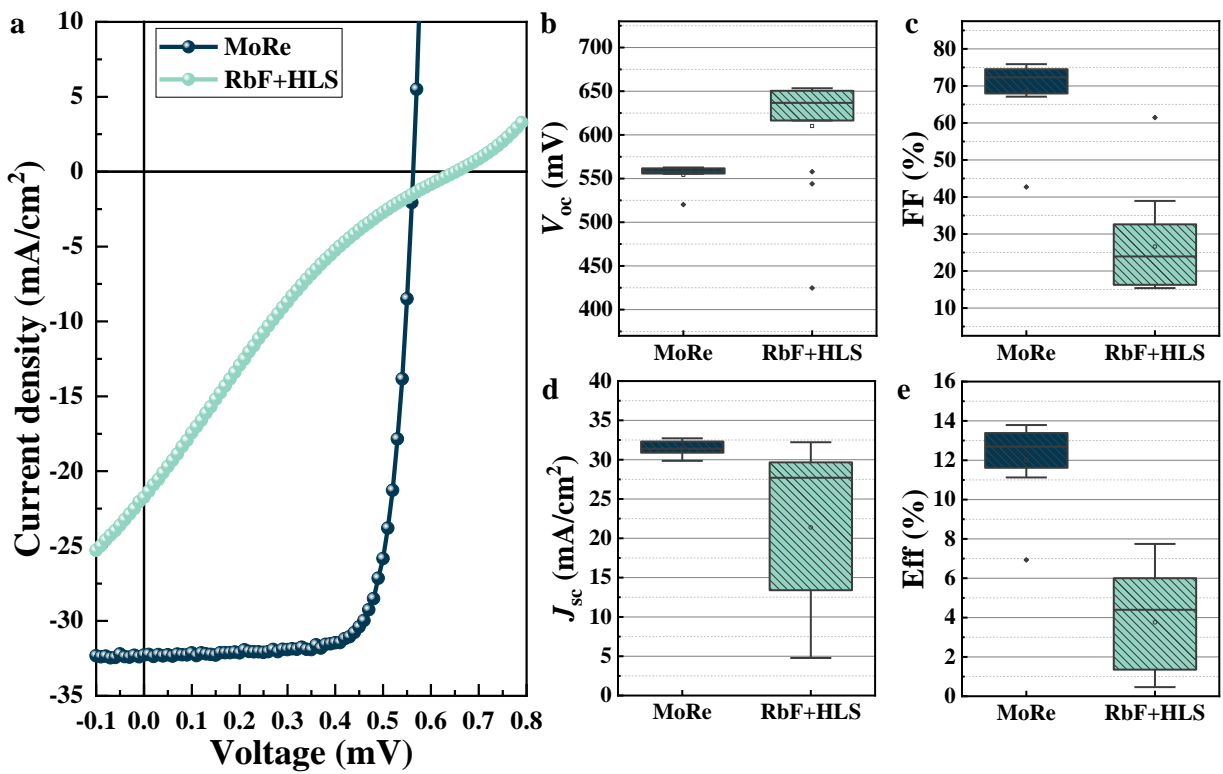

**Extended Date Fig. 6| a,** The *J-V* curves depict the performance of both the reference solar cell and the HTL passivated solar cells with RbF PDT and heat light soaking. The $In_2O_3$ layer of the HTL is fabricated using sputter with thicknesses of 30 nm. The *J-V* parameters of the solar cells are presented as follows: **b,** $V_{oc}$; **c,** FF; **d,** $J_{sc}$, and **e,** Eff.

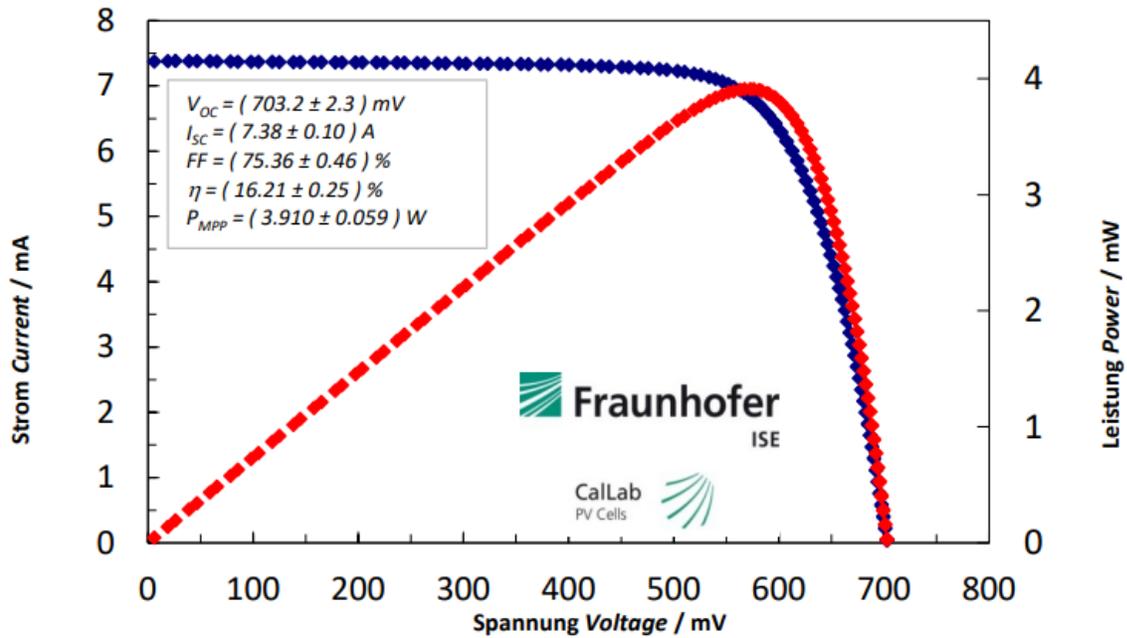

**Extended Date Fig. 7|** Certified *J-V* measurement from ISE CalLab.